\documentclass[preprint,eqsecnum,aps,psfig,epsfig,citesort]{revtex}
\newif\iftightenlines\tightenlinesfalse
\tightenlines\tightenlinestrue
\begin{document}

%%%%%%%%%%%%%%%%
%
%%%%%%%%%%%%%%%%%
\begin{titlepage}
%%%%%%%%%%%%%%%%%
%
\begin{flushright}
BNL-NT-03/12 \\
RBRC-324 \\
June 2003
\end{flushright}
\begin{center}
\large
{\bf {Target Mass Corrections to Electro-Weak Structure Functions}}\\
\medskip
{\bf {and}}\\
\medskip
{\bf {Perturbative Neutrino Cross Sections}}\\
\vspace*{0.5cm}
\large 
{S.\ Kretzer$^{a,b}$ and M.H.\ Reno$^{c}$}
\vspace*{0.5cm}
\normalsize

{\em $^a$Physics Department, Brookhaven National Laboratory,\\
Upton, New York 11973, U.S.A.}\\

%\vspace*{0.5cm}
{\em $^b$RIKEN-BNL Research Center, Bldg. 510a, Brookhaven 
National Laboratory, \\
Upton, New York 11973 -- 5000, U.S.A.}

%\vspace*{0.5cm}
{\em $^c$ Department of Physics and Astronomy, University of Iowa \\
Iowa City, Iowa 52242 USA}
\end{center}
%\vspace*{0.5cm}
%%%%%%%%%%%%%%%%
\begin{abstract}
%%%%%%%%%%%%%%%%                                                            
We provide a complete and consistent framework to include subasymptotic perturbative
as well as mass corrections to the leading twist ($\tau=2$)
evaluation of charged and neutral current weak structure 
functions and the perturbative neutrino cross sections. We revisit previous calculations
in a modern language and fill in the gaps that we find missing for a complete 
and ready-to-use ``NLO $\xi$-scaling" formulary. In particular, as a new result we 
formulate the mixing of the partonic and hadronic structure function tensor basis 
in the operator approach to deep inelastic scattering. 
As an underlying framework we follow the operator 
product expansion \`{a} la Georgi \& Politzer that 
allows the inclusion of target mass corrections 
at arbitrary order in QCD and we provide explicit
analytical and numerical results at NLO. We compare this approach with a 
simpler collinear parton model approach to $\xi$-scaling. 
Along with target mass corrections we include 
heavy quark mass effects as a calculable leading twist power suppressed correction. 
The complete corrections have been implemented into a Monte Carlo integration program to
evaluate structure functions and/or integrated cross sections.
As applications, we compare the operator approach with the collinear approximation numerically
and we investigate the NLO and mass corrections to observables that are related to 
the extraction of the weak mixing angle from a Paschos-Wolfenstein-like relation in
neutrino-iron scattering. We expect that 
the interpretation of neutrino scattering events in terms of oscillation 
physics and electroweak precision
physics will benefit from our results.     
\end{abstract}
\end{titlepage}
\newpage
%%%%%%%%%%%%%%%%%%%%%%%%%%%%%%%%%%%%%%%%%%%%%%%%%%%%%%%%%%%%%%%%%%%%%%%%%
\section{Introduction}
Recent neutrino experiments have shown strong evidence for neutrino
masses and mixing. A variety of 
experiments \cite{Fukuda:1998mi,Ahmad:2002ka} have
yielded data that when combined \cite{Bahcall:2002hv,Maltoni:2002ni} 
have led to a picture
of neutrino mass and mixing with large mixings
between $\nu_e\leftrightarrow\nu_\mu$ and $\nu_\mu\leftrightarrow
\nu_\tau$, $\Delta m_{12}^2\sim 7\times 10^{-5}$ 
eV$^2$ and $\Delta m_{23}^2\sim
2\times 10^{-3}$ eV$^2$. Solar neutrinos are produced with $E_\nu < 20$ 
MeV \cite{Bahcallbook},
making $\nu_e\rightarrow \nu_\mu$ the relevant oscillation process.
For the atmospheric case, $\langle E_\nu\rangle\sim 1$ GeV and
$\nu_\mu\rightarrow \nu_\tau$ is the dominant oscillation process. 
The MINOS, OPERA and Chorus experiments will study 
$\nu_\mu\rightarrow \nu_\tau$ oscillations, either by $\nu_\mu$
disappearance or $\nu_\tau$ appearance \cite{exps}.  

In the GeV neutrino energy range, target mass effects are
important. For muon neutrino charged current interactions,
at low energies ($E_\nu\sim 100-800$ MeV), the
quasi-elastic process dominates and nucleon mass effects are easy
to incorporate \cite{LlewellynSmith:1971zm}.  At intermediate energies,
exclusive few pion production processes are the largest
contributions to the cross section 
\cite{Lipari:1994pz,Paschos:2001np,gallagher}.
Above a few GeV, deep inelastic
scattering (DIS) dominates the cross section. 
The energies for tau neutrino charged current interactions are higher
because of the tau lepton threshold. 
A series of conferences ({\it NUINT}) has grown out of the efforts
to improve and combine knowledge about the distinct neutrino interaction modes
into event generation for oscillation searches \cite{nuint}.

In a different branch of recent research at the interface of neutrino
physics and QCD, 
the NuTeV collaboration has measured the weak mixing angle
from charged and neutral current muon neutrino and anti-neutrino interactions 
with a broad band beam (see Fig.~\ref{fig:flux}).  
The beam flux maximum is at an energy of 
about $E_{\nu}\simeq 60\ {\rm GeV}$ which translates
into an average event energy of about $E_{\nu}\simeq 100\ {\rm GeV}$
because the cross section rises approximately linearly with energy. 
NuTeV extracts a value of $\sin^2 \Theta_{\rm W}=0.2277\pm 0.0013 \pm 0.0009$ 
\cite{nuanom} which lies about $3 \sigma$ above the standard model value 
$0.2227  \pm 0.00037$ \cite{lepew}. 
The latter fit value does not  
include neutrino scattering data. Possible explanations of the 
{\it NuTeV anomaly} within and beyond the standard model have been  
mapped out in Ref.~\cite{forte}. Nuclear physics related complications 
in its interpretation - the NuTeV target material consists mostly of steel -
have also been discussed in Ref.~\cite{nuref}. No consistent picture has as yet
emerged from these efforts to understand the measured value.   

The perturbative QCD parton picture is applicable at NuTeV energies and at 
the higher energy end of the oscillation neutrino beams. A perturbative background to
(quasi-)elastic scattering is present at lower energies as well. 
At none of these energies can sub-asymptotic corrections to the simplest 
parton model picture be neglected, and both oscillation
and precision neutrino physics are bound to benefit from a concise inclusion
of these terms into data analysis. At least for precision physics, it does 
not suffice to model these corrections but it is a necessity to extract
theory parameters using theoretically sound cross sections. E.g.~a well defined
separation of higher twist effects (see e.g.~\cite{sss})
requires that power suppressed mass terms
are accounted for correctly in the leading twist-2 cross section. 

In the present article we include
\begin{itemize}
\item ${\cal{O}}(\alpha_s )$ perturbative QCD corrections
\item ${\cal{O}}(M^n/Q^n)$ target mass effects 
\item ${\cal{O}}(m^n/Q^n)$ 
and ${\cal{O}}(\ln m^2/Q^2)$ heavy quark mass effects
\item ${\cal{O}}(m_l^{2n}/(M^n E_l^n))$ heavy lepton mass effects
(mostly for $\tau$ production) 
\end{itemize}
and any combination of the above. One could summarize the above as
(Nachtmann or Georgi \& Politzer) $\xi$-scaling for weak structure
functions merged with NLO QCD for light and heavy quarks: While the heavy quark
and perturbative corrections come in through the expansion of the Wilson
coefficients in the operator product expansion, the target mass corrections find their
way into the final result through the Lorentz structure of the corresponding (non-reduced)
operators. 

Target mass corrections (TMC)
to DIS have been know for a long time. The first discussion in terms
of the operator product expansion (OPE) at leading order in QCD
was done by Georgi and 
Politzer \cite{gp} in 1976. Later, these same target mass corrections
were derived from a parton model point of view by Ellis, Furmanski and
Petronzio \cite{efp}. DeRujula, Georgi and Politzer discussed
next-to-leading order (NLO) QCD corrections to target mass corrected
structure functions $W_{1,2}$
in the context of local duality in electroproduction \cite{dgp} 
and using off-shell regularization. In this paper, we present the
target mass corrections for charged current (CC) and weak neutral current (NC)
$\nu_\mu$ and $\nu_\tau$ DIS with
nucleons, including next-to-leading order QCD corrections 
and heavy quark production  using modern conventions. 

In the next Section, we outline the procedure for evaluating the TMC
beyond leading order in QCD within the OPE. In Section \ref{sec:res}, 
we show our results in analytic form and compare them to the collinear
parton model. We also make a few comments on nuclear targets and on the
validity of the Albright \& Jarlskog relations for the structure functions
$F_{4,5}$. Numerical results
for the structure functions are shown in Section \ref{sec:appl} for both the 
OPE and parton model approach to target mass corrections. As a further 
application, we evaluate observables that are related to the NuTeV Weinberg angle
analysis in NLO QCD with high numerical precision. 
A summary appears in Section \ref{sec:sum}.

\section{Target Mass Corrections}
\label{sec:tmc}

\subsection{Parton Model Approach (Briefly Revisited)}

Target mass corrections appear in a number of ways that can be described
in parton model language. First, the parton fraction $\xi$ (Nachtmann 
variable \cite{nacht})
of the light cone momentum of the nucleon $P_N^+=(P_N^0+P_N^z)/\sqrt{2}$
is related to the Bjorken $x$ by
\begin{equation}
{1\over \xi}={1\over 2x}+\sqrt{{1\over 4x^2}+{M^2\over Q^2}}
\ \Longleftrightarrow
\ \xi = \frac{2 x}{1+\sqrt{1+\frac{4 M^2 x^2}{Q^2}}} 
\end{equation}
for massless quarks. Second, there is a mixing between partonic structure
functions in the evaluation of the hadronic structure functions.
The partonic light cone momentum is related to the
hadronic light cone momentum via $p^+=\xi P_N^+$, however,
partonic momentum projections are not simple rescalings
of the hadronic momentum projections on the hadronic
tensor $W^{\mu \nu}$ because $p^- \neq \xi P_N^-$ when 
$\{p^2,P_N^2\}=\{0,M^2\}$. Finally, in a collinear expansion 
\cite{efp} target mass effects appear from transverse 
momentum ($k_T$) effects. 

In the collinear
parton model ($k_T =0$ approximation), 
target mass corrections have been evaluated
in Ref.~\cite{aot} and later in Ref.~\cite{kr}. 
The intrinsic $k_T$ of the target 
partons is limited by the nucleon mass $M$, which introduces further
corrections. 
Ellis, Furmanski and Petronzio \cite{efp}
showed the equivalence of a non-collinear parton approach results,
where the parton is on-shell but not collinear with the nucleon, 
to the operator product expansion target mass
corrections, to which we will turn next.

\subsection{Operator Product Expansion}

The three contributions of the partonic
approach appear automatically in the operator product
expansion approach to target mass corrections, first discussed in this
context by Georgi and Politzer \cite{gp}. 
In this approach, one starts with the
relation between the hadronic tensor $W^{\mu\nu}$ and the virtual
forward Compton scattering amplitude $T^{\mu\nu}$. 
For weak interactions,
the relevant set of structure functions is $W_i,\ i=1-5$:
\begin{eqnarray} \label{eq:wmunu0} 
W_{\mu\nu} && \equiv
{1\over {2 \pi}}\ \int e^{iq\cdot z}d^4 z 
\langle N|\left[J_\mu (z), J_\nu (0)\right]
|N\rangle
\\ 
\label{eq:wmunu2}
&& = {1\over {\pi}}\ {\rm Disc}\ \int e^{iq\cdot z}d^4 z 
\langle N|iT \left(J_\mu (z) J_\nu (0)\right)
|N\rangle \\
\label{eq:wmunu3}
&& \equiv {1\over {\pi}}\ {\rm Disc}\ T_{\mu\nu}\\
\label{eq:wmunu1}
&&=-g_{\mu \nu}W_1 + {p_\mu p_\nu\over M^2}W_2
-i\epsilon_{\mu\nu\rho\sigma}
{p^{\rho}q^\sigma\over M^2}W_3
+{q_\mu q_\nu \over M^2}W_4+{p_\mu q_\nu+p_\nu q_\mu\over M^2} W_5\ .
\end{eqnarray}
In Eqs.~(\ref{eq:wmunu2}), (\ref{eq:wmunu3}) the discontinuity
is taken across the $\nu=q^0>0$ positive frequency cut:
\begin{equation} \label{eq:cut}
{\rm Disc}\ T_{\mu\nu} \equiv \lim_{\varepsilon \rightarrow 0}
\ \frac{1}{2 i}\ \left[ 
T_{\mu\nu}(\nu+i \varepsilon ) - T_{\mu\nu}(\nu-i \varepsilon )
\right] \ .
\end{equation} 
The hadronic tensor $W_{\mu\nu}$ is probed in deep inelastic interactions
with a leptonic current 
\begin{equation}\label{eq:jlep}
J_{\mu}^l = {\overline \psi_2} \gamma_{\mu} (v_l - a_l \gamma_5 ) \psi_1
\end{equation}
through exchange of a boson of virtuality $Q^2$. The corresponding cross
section in the target rest frame 
is
\begin{equation}\label{eq:sigdis}
\frac{d^2\sigma^{\nu(\bar{\nu})}}{dx\ dy} = \frac{G_F^2 M
E_{\nu}}{\pi(1+Q^2/M_B^2)^2}\
\sum_{i=1}^{5}
w_i(x,y,E_{\nu},v_l,a_l,m_1,m_2)
F_i (x,Q^2) 
\end{equation}
where $M_{B=W^{\pm}, Z^0}$ is the boson mass,
$x$ and $y$ are the standard DIS variables 
and
where the modern normalizations are
\begin{equation}\label{eq:WtoF}
\left\{F_1,\ F_2,\ F_3,\ F_4,\ F_5\right\}\ = 
\ \left\{W_1, 
\ \frac{Q^2}{2 x M^2}W_2 ,
\ \frac{Q^2}{x M^2}W_3 ,
\ \frac{Q^2}{2 M^2} W_4 ,
\ \frac{Q^2}{2 x M^2} W_5
\right\}\ . 
\end{equation}
For purely electromagnetic interactions\footnote{Interference contributions 
between $\gamma^\ast$ and $Z^0$
at very high $Q^2$ (e.g.~at HERA) are not sensitive to target mass 
corrections. They are, therefore, not covered by 
Eqs.~(\ref{eq:sigdis}), (\ref{eq:sigdisem}).}
we have
\begin{equation}\label{eq:sigdisem}
\frac{d^2\sigma^{l^{\pm}}}{dx\ dy} = \frac{16 \pi \alpha^2  M
E_{l}}{ Q^4}\
\sum_{i=1}^{2}
w_i(x,y,E_l,1,0,m_l,m_l)
F_i (x,Q^2) \ .
\end{equation}
The electroweak weights $w_i$ in 
(\ref{eq:sigdis}), (\ref{eq:sigdisem}) 
are listed in Eqs.~(\ref{eq:wa1})-(\ref{eq:wc5}) 
for general lepton masses and couplings
along with the experimentally relevant cases. 

As is well known, $W_3$ contributes only to weak processes because
of the parity violating nature of the anti-symmetric
$\varepsilon$-tensor. A sixth independent
tensor combination $\sim W_6 ( p_\mu q_\nu-p_\nu q_\mu )$ 
does not couple to the the most general leptonic current
\begin{equation} \label{eq:w6}
L^{\mu \nu}(v_l,a_l,m_1,m_2)  \left( p_\mu q_\nu-p_\nu q_\mu \right)=0
\ \ \forall \{v_l,a_l,m_1,m_2\}
\end{equation}
where $ L^{\mu \nu}$ is the
leptonic analogue of the hadronic $W_{\mu\nu}$. 
Because $W_6$ decouples from the DIS process, we 
omit it from the beginning and throughout.

In neutral current reactions, $W_{4,5}$ contribute through the axial 
coupling $a_l$ only, i.e.~they do not couple to a $\gamma^\ast$ exchange.
In standard formulae, the tensors corresponding to $W_4$ and $W_5$ 
are sometimes combined with those corresponding to 
to $W_1$ and $W_2$ so that current conservation is manifest in the
electromagnetic case.
Generally, for both neutral and charged current weak reactions
the structure functions $W_4$ and $W_5$
do not appear in the usual discussion since they are multiplied by
the ratio of the lepton mass squared to $M E_\nu$. 
They obey the Albright-Jarlskog relations \cite{aj} which are valid
at leading order in the limit of massless quarks and a 
massless target\footnote{The conditions under which Eqs.~(\ref{eq:aj1})
and (\ref{eq:aj2}) are valid individually are detailed in the appendix
of Ref.~\cite{kr} and in Section \ref{sec:ml} below.
}: 
\begin{eqnarray} \label{eq:aj1}
F_4 & & = 0\\
2x F_5&& =F_2 \label{eq:aj2}
\end{eqnarray}

The forward scattering amplitude, explicitly including
perturbative corrections, can be written as
\begin{eqnarray} \label{eq:tmunu}
T^{\mu\nu} &&= \sum_{k=1}^\infty\Bigl( -g^{\mu\nu} q_{\mu_1}q_{\mu_2} C_1^{2k}
+ g^\mu_{\mu_1}g^\nu_{\mu_2} Q^2 C_2^{2k} - i\epsilon ^{\mu\nu\alpha\beta}
g_{\alpha\mu_1}q_\beta q_{\mu_2}C_3^{2k}\\ \nonumber
&&+{q^\mu q^\nu\over Q^2}q_{\mu_1}q_{\mu_2} C_4^{2k}
+( g^\mu_{\mu_1}q^\nu q_{\mu_2} + g^\nu_{\mu_1}q^\mu q_{\mu_2} )C_5^{2k}
\Bigr) q_{\mu_3}...q_{\mu_{2k}}{2^{2k}\over Q^{4k}} A_{2k} \Pi^{\mu_1 ...
\mu_{2k}}\ ,
\end{eqnarray}
where $A_{2k}$ represents the reduced (scalar) matrix element of a twist-2 
(traceless and symmetric) operator 
${\cal{O}}_{\tau =2}^{\mu_1 ... \mu_{2k}}$
of spin $2 k$ 
\begin{equation} \label{eq:redmat}
\left< N \right| {\cal{O}}_{\tau =2}^{\mu_1 ... \mu_{2k}} \left| N \right>
= A_{2k} \Pi^{\mu_1 ... \mu_{2k}} \ .
\end{equation}
The QCD perturbative corrections come in through non-trivial 
Wilson coefficients $C_i^{2k}$. 
In Eq.~(\ref{eq:tmunu}), the factorization
into universal operator matrix elements and process-dependent coefficient 
functions is a consequence of the operator product expansion\footnote{ 
The OPE has found its way into many excellent textbooks and 
clear pedagogical introductions can be found in e.g.~\cite{textbooks}.}
\cite{ope}
of the product of currents $T^{\mu\nu}$ in Eq.~(\ref{eq:wmunu2}).
We assume that ${\cal{O}}_{\tau =2}^{\mu_1 ... \mu_{2k}}$ 
is of twist 2 to make contact with the parton model. 
For the following evaluation of the target mass corrections, however,
it is sufficient that ${\cal{O}}$ is a traceless and symmetric operator.
Symmetry under permutation
and tracelessness under contraction of any pair of indices of 
${\cal{O}}_{\tau =2}^{\mu_1 ... \mu_{2k}}$
require then 
that the tensor part of the matrix element in 
Eq.~(\ref{eq:redmat}) is
\begin{equation}\label{eq:pi}
\Pi^{\mu_1 ...\mu_{2k}}=\sum_{j=0}^k (-1)^j {(2k-j)!\over 2^j (2k)!}
\underbrace{g\ .\ .\ .\ g}_{j\ {g^{\mu_i\mu_j}}{\rm 's}} \ \underbrace{p\ .\ .\ .\ p}_
{(2k-2j)\ {p^{\mu_i}}{\rm 's}}\ (p^2)^j\ ,
\end{equation}
where $g\ ...\ g\ p\ ...\ p$ abbreviates a sum over $(2k)! / [2^j j!(2k-2j)!]$
permutations (not counting identical terms such as $g^{\mu_i\mu_j} =
g^{\mu_j\mu_i}$ twice) of the indices.
In Eq.~(\ref{eq:pi}) the $j=0$ approximation $\Pi^{\mu_1 ...\mu_{2k}} \propto
p^{\mu_1} .... p^{\mu_{2k}}$ reproduces the massless parton model 
and the ${j>0}$ terms resum the target mass corrections. 
While they are formally power suppressed of order ${\cal{O}}(M^{2k}/Q^{2k})$,
the corrections derive from the same operators
${\cal{O}}_{\tau =2}^{\mu_1 ... \mu_{2k}}$ as the $j=0$ term
and as such are of twist-2. This is in contrast to power corrections from operators
of higher twist ${\cal{O}}_{\tau > 2}^{\mu_1 ... \mu_{2k}}$ which introduce
new non-perturbative matrix elements 
independent of those in Eq.~(\ref{eq:redmat}).
In classical terminology twist classifies strictly the dimension 
minus the spin of an operator; the difference between target mass
corrections and higher twists is, nevertheless, sometimes referred to 
as a distinction between kinematic and dynamical higher twists as well.

The quantity $A_{2k}$ is the $2k^{\rm th}$ moment of a universal function $f$ 
or the $(2k-1)^{\rm th}$ moment of the
quark distributions $q(x)$\footnote{For ease of notation,
throughout this article we sometimes 
suppress scale dependence as a functional argument.
$Q^2$-dependence enters through the renormalization of 
${\cal{O}}_{\tau =2}^{\mu_1 ... \mu_{2k}}$; i.e.~strictly $A_{2k}=A_{2k}(Q^2)$. 
Also, for the purpose of this article and in order to make contact with 
the parton model it suffices to assume that
${\cal{O}}_{\tau =2}^{\mu_1 ... \mu_{2k}}$ 
is a quark operator since the inclusion of
operators built from gluons and operator mixing would not 
alter the discussion anywhere.
}
\begin{equation} \label{eq:a2k}
A_{2k} = \int_0^1 dx\ x^{2k} f(x,Q^2) = \int_0^1 dx\ x^{2k-1} q(x,Q^2)\ .
\end{equation}
Beyond leading order we include the non-universal Wilson coefficients
into the moment integral
\begin{equation}\label{eq:c2ka2k}
C_i^{2k}A_{2k} = \int_0^1 dx\ x^{2k}\ f_i(x,Q^2)\ ,
\end{equation}
thus defining a set of non-universal 
$x$-space functions $f_{i=1,...5}$. The 
$f_i$ can be related to the zero mass limit $F_i^{(0)}$ of 
the experimental structure functions
$F_i$ from the $j=0$ term of the series (\ref{eq:pi}) as will
be explained in the next section; the effect of the
resummation of the $j>0$ terms will induce convolution integrals
over the $F_i^{(0)}$ and lead to Nachtmann scaling \cite{nacht}.

The standard calculation \`a la Georgi-Politzer of extracting the moments
of the structure functions, then performing inverse Mellin transforms,
leads to equations most simply expressed in terms of $f_i(\xi,Q^2)$ 
and
\begin{eqnarray} \label{eq:hi}
h_i(\xi,Q^2)&&=\int_\xi^1 d \xi' \ f_i(\xi',Q^2)\\ \label{eq:gi}
g_i( \xi,Q^2)&&=\int_\xi^1 d \xi' \ h_i(\xi',Q^2)=\int_\xi^1 d\xi' (\xi'-\xi)
f_i(\xi',Q^2) .
\end{eqnarray}
The middle part of Eq.~(\ref{eq:gi}) is a more transparent representation
when we write down the results below. For numerical evaluation, however, it
is more convenient to express the double convolution integral as a moment
of a single convolution as suggested by the right-hand-side of the same 
equation.

We have extended the Georgi-Politzer \cite{gp}  and 
DeRujula-Georgi-Politzer \cite{dgp}  analyses 
by including the full
set of weak structure functions along with their Wilson coefficients
$C_i^{2k}$. 
Abelian gauge invariance is not
assumed through {\it a priori} relations among $C_i^{2k}$, although it
is restored in the electromagnetic scattering case by the explicit
values of $C_i^{2k}$. 
In unpolarized DIS, the leading component is of twist-2 for any
$W_{i=1,...,5}$, and the coefficient functions do the bookkeeping
of the tensor components in $W_{\mu \nu}$.
One finds a mixing of structure functions,
which for $F_1$ was already found in Ref.~\cite{dgp}.
The general pattern of mixing can be easily inferred from
the $j>0$ terms in Eq.~(\ref{eq:pi}) that add different tensor components
to the $j=0$ term.
We note that in polarized DIS, the spin dependent structure
functions \cite{wandzura,mu,pr,bt} $g_1$ and $g_2$ receive contributions
from twist-2 and twist-2 and 3 operators, respectively.
In this case, tracking the Wilson coefficient basis 
follows automatically from tracking the operators of twist two and three
and from the Wandzura-Wilczek relation for 
the twist-2 part \cite{wawi}.

The method of evaluating target mass corrections as inverse Mellin
transforms of moments of structure functions was presented in
explicit detail in the Georgi \& Politzer analysis \cite{gp} already.
Recent analogous calculations in the spin dependent case 
\cite{pr,bt} have repeated the mathematical steps involved. 
Apart from the differences listed above, we feel it is unnecessary to 
add a further derivation to the literature. 
In the next section, we show our 
explicit results for the target mass corrected structure
functions $W_i$ and the corresponding $F_i$.

\section{Results}
\label{sec:res}

\subsection{Generic Formulae}

The procedure outlined above
leads directly
to the results for the structure functions in terms of Bjorken $x$,
the Nachtmann variable $\xi$, and $Q^2$ as well as $\mu\equiv M^2/Q^2$.
The structure functions are:
\begin{eqnarray} \label{eq:w1tmc}
W_1 &&={1\over 2}\ {\xi\over 1+\mu \xi^2}f_1(\xi,Q^2)-\mu x^2{\partial\over
\partial x}\Bigl[ {g_2(\xi,Q^2)\over 1+\mu\xi^2}\Bigr]\\ \label{eq:w2tmc}
W_2 &&= {2\mu}x^3{\partial^2\over \partial x^2}\Bigl[{xg_2(\xi,Q^2 )\over
\xi (1+\mu\xi^2 )}\Bigr]\\ \label{eq:w3tmc}
W_3 &&=-\mu x^2{\partial\over \partial x}
\Bigl[ {h_3(\xi,Q^2)\over (1+\mu \xi^2)}\Bigr] \\ \label{eq:w4tmc}
W_4 &&={\mu \xi\over 2 (1+\mu\xi^2)}f_4(\xi,Q^2)+2\mu^2 x^2{\partial\over
\partial x}\Bigl[ {\xi h_5(\xi,Q^2 )\over 1+\mu \xi^2}\Bigr]
+ 2\mu^3 x^3{\partial ^2\over \partial x^2}\Bigl[ {\xi^2g_2(\xi,Q^2)\over
(1-\mu\xi^2)(1+\mu \xi^2)}\Bigr]\\ \label{eq:w5tmc}
W_5 & &  = -\mu x^2{\partial\over \partial x}
\Bigl[ {h_5(\xi,Q^2)\over (1+\mu \xi^2)}\Bigr]
-2\mu^2 x^3{\partial^2\over \partial x^2}\Bigl[{\xi g_2(\xi,Q^2)
\over (1-\mu \xi^2)
(1+\mu \xi^2)}\Bigr]
\end{eqnarray}
After performing the derivatives in Eqs.~(\ref{eq:w1tmc})-(\ref{eq:w5tmc}), 
one gets a generic form
\begin{equation}
W_j =\sum_{i=1,5} a_j^i f_i(\xi,Q^2) + b_j^i h_i(\xi,Q^2) + c_j g_2(\xi,Q^2)\ .
\end{equation}

The generic formula for the $W_j$ applicable to Eq.~(\ref{eq:wmunu1}) is more
useful translated to modern normalization conventions of the
structure functions ($F_j$) or to the parton model language.
In the next sections, we associate the $f_i$ and related $h_i$ and
$g_2$ to the experimental structure functions $F_i(x,Q^2)$ in the
$M\rightarrow 0$ limit. The $\left. F_i(x,Q^2)\right|_{M=0}$
can be evaluated at NLO using parton distribution functions
as documented in detail in \cite{nlo,kr}. As in \cite{kr}
and as is appropriate for the neutrino energies 
under consideration, 
we will be working in the
${\overline{\rm MS}}$ factorization scheme with $n_f=3$ 
active flavours. 
To avoid redundancy,
we will not reproduce the formulae of \cite{nlo,kr} in the present 
article.
In Section \ref{sec:nucl} we comment on nuclear
target corrections. We then consider the association  of
$W_j$ to parton
distribution functions in Section \ref{sec:coll}. The master formulae for $F_j$ in
structure function language and in parton model language appear in
Eqs.~(\ref{eq:FTMC}) and (\ref{eq:FcalTMC}) below.

\subsection{Target Mass Corrected Structure Functions}

Our procedure for determining the quantities $f_i$ and associated $h_i$ and 
$g_i$ is to consider the experimental structure functions $F_i$
in Eq.~(\ref{eq:WtoF})
with the $W_i$ in Eqs.~(\ref{eq:w1tmc})-(\ref{eq:w5tmc})
in the $\mu\rightarrow 0$ limit 
\begin{equation}
{F_i}^{(0)} \equiv \left.{F_i}\right|_{M=0}\ ,
\end{equation}
in which $\xi\rightarrow x$. Note that the ${F_i}^{(0)}$ here 
denote functional forms and not observables -- though they are accessible 
in principle through the inversion of Eq.~(\ref{eq:FTMC}) below as the
structure functions with moments of definite 
spin in the Nachtmann sense \cite{nacht}
or the perturbatively corrected scaling functions 
in the duality sense \cite{dgp}.
To make the distinction from ${F_i}^{(0)}$
more obvious we will, below,
sometimes label the measurable structure functions ${F_i}^{\rm TMC}$. 
From the $j=0$ term of the target mass series in Eq.~(\ref{eq:pi}) we 
find the following relations between $f_i$ and the zero-target-mass
limit functions ${F_i}^{(0)}$
\begin{equation} \label{eq:wtof}
\left\{F_1^{(0)}(x), F_2^{(0)}(x), F_3^{(0)}(x), F_4^{(0)}(x), F_5^{(0)}(x)\right\}= 
\left\{\frac{x}{2} f_1(x), 
x^2f_2(x) ,
x f_3(x) ,
\frac{x}{4} f_4(x) ,
\frac{x}{2} f_5(x)
\right\} \ . 
\end{equation}
In Section III.C, we show that this form explicitly substantiates
the partonic interpretation 
\begin{equation} \label{eq:ftoq}
f_i(x) = k_i\ f(x) + {\cal{O}}\left( \alpha_s \right) =
k_i \ \frac{q(x)}{x} + {\cal{O}}\left( \alpha_s \right)
\end{equation}
with the couplings of the quark current
\begin{equation}\label{eq:jquark}
J_{\mu}^q = {\overline \psi_2} \gamma_{\mu} (v_q - a_q \gamma_5 ) \psi_1
\end{equation}
entering as\\ \begin{center}
\begin{tabular}{ll}
$k_i = v_q^2 + a_q^2$ & (i=1,2,5)\\
$k_i = 2\ v_q a_q$ & (i=3)\\
$k_i = 0$ & (i=4)\ .
\end{tabular}\\ \end{center}
The form of Eq.~(\ref{eq:ftoq}) is modified slightly to include quark
masses.
For the applications in Section \ref{sec:appl} we calculate the $F_j^{(0)}$
using the ${\cal{O}}(\alpha_s )$
coefficient functions in Refs.~\cite{nlo,kr}.
Here we focus on relating $F_i^{TMC}$ to the $F_j^{(0)}$
and their convolutions.
With Eq.~(\ref{eq:wtof}), Eqs.~(\ref{eq:hi}) and (\ref{eq:gi}) now become
\begin{eqnarray} \label{eq:h1}
h_1(\xi,Q^2)&&=\int_\xi^1 d \xi' \ \frac{2 F_1^{(0)}(\xi',Q^2)}{\xi'} \\
\label{eq:2}
h_2(\xi,Q^2)&&=\int_\xi^1 d \xi' \ \frac{  F_2^{(0)}(\xi',Q^2)}{{\xi'}^2} \\
\label{eq:h3}
h_3(\xi,Q^2)&&=\int_\xi^1 d \xi' \ \frac{  F_3^{(0)}(\xi',Q^2)}{\xi'} \\
\label{eq:h4}
h_4(\xi,Q^2)&&=\int_\xi^1 d \xi' \ \frac{4 F_4^{(0)}(\xi',Q^2)}{\xi'} \\
\label{eq:h5}
h_5(\xi,Q^2)&&=\int_\xi^1 d \xi' \ \frac{2 F_5^{(0)}(\xi',Q^2)}{\xi'}
\\ 
\label{eq:g2} 
g_2( \xi,Q^2)&&=\int_\xi^1 d \xi' \ h_2(\xi',Q^2)
=\int_\xi^1 d\xi' (\xi'-\xi)\ \frac{  F_2^{(0)}(\xi',Q^2)}{{\xi'}^2} 
\end{eqnarray}
From Eqs.~(\ref{eq:w1tmc})-(\ref{eq:w5tmc}), (\ref{eq:wtof}), (\ref{eq:WtoF}), 
(\ref{eq:h1})-(\ref{eq:g2}), 
the target mass corrected structure functions then are
\begin{equation}\label{eq:FTMC}
F_j^{\rm TMC}=\sum_{i=1,5} A_j^i {F}_i^{(0)}(\xi,Q^2) + B_j^i 
{h}_i(\xi,Q^2) + C_j {g}_2(\xi,Q^2)\ .
\end{equation}
The coefficients $\{A_j^i,B_j^i,C_j\}$ are listed 
in Table \ref{tab:A}-\ref{tab:C} in terms
of $x$, $\xi$, $\mu$ and $\rho$ where
\begin{equation}
\rho\equiv {1+\mu\xi^2\over 1-\mu\xi^2}= \sqrt{1+4\mu x^2}\ .
\end{equation}
Note that 
\begin{equation}
x={\xi\over 1-\mu \xi^2}\ .
\end{equation}
While we have sometimes referred to twist-2 operators to meet the parton model,
Eq.~({\ref{eq:FTMC}) holds for contributions from any traceless 
and symmetric operators and it is independent of the perturbative order in $\alpha_s$.
For leading order ${\cal{O}}(\alpha_s^0)$ twist-2 coefficient functions,
$F_{1,2}$ as in Eq.~({\ref{eq:FTMC}) agree with the standard results
in Ref.~\cite{gp} and the non-collinear parton results in Ref.~\cite{efp}. 
For $F_3$ there is a factor of $2 x$ mismatch in relative 
normalization of the $F_3^{(0)}$ and $h_3$ term between Ref.~\cite{gp} and 
Ref.~\cite{efp}; we agree with the results in Ref.~\cite{efp}.

\subsection{A Note on Nuclear Targets}\label{sec:nucl}
By the very nature of weak interactions heavy nuclei are more natural
targets than a nucleon, so a brief comment on large $M_A=A M_N$ nuclei may be
worthwhile. 
In the OPE the coefficient functions are manifestly
target independent and 
the target dependence resides in a generalization of 
Eq.~(\ref{eq:redmat}) with $\left|N\right>$ replaced by a general nuclear
target
\begin{equation}\label{eq:redmatA} 
\left< Z,A \right| {\cal{O}}_{\tau =2}^{\mu_1 ... \mu_{2k}} \left| Z,A \right>
= A_{2k}^{(Z,A)} \Pi^{\mu_1 ... \mu_{2k}} \ .
\end{equation}
i.e.~ultimately the nuclear dependence for $\tau=2$ is contained 
in nuclear modifications of the PDF in Eq.~(\ref{eq:a2k})\footnote{
In the absence of nuclear effects
we could drop the label
``$Z,A$" on $q^{(Z,A)}$ in Eq.~(\ref{eq:a2kA}).}
}
\begin{equation} \label{eq:a2kA}
A_{2k}^{(Z,A)} = A \int_0^1 dx\ x^{2k-1} q^{(Z,A)}(x,Q^2)\ \ .
\end{equation}
Genuine higher twist $\tau>2$ effects
may be of different relative size in nuclei, 
e.g.~the twists investigated in Ref.~\cite{onethird} are enhanced by a factor
$A^{1/3}$. 
The $M_A^2=A^2 M_N^2$ enhancement of target mass
terms is, however, balanced everywhere by kinematic factors that scale 
with inverse powers of $A$ such as
\begin{equation}
x_A = \frac{Q^2}{2 P_A \cdot q} = \frac{Q^2}{2 A P_N \cdot q}=\frac{x}{A}
\end{equation}
Neglecting for the moment the nuclear modifications in 
Eq.~(\ref{eq:a2kA}),
under the assumption that\footnote{The isospin orientation
is attributed to the PDF without explicit labeling.}
\begin{equation}\label{eq:FA}
q^{(Z,A)}(x)=A\ q(x)\ \Theta (1-x)
\end{equation}
target mass corrections are then the same for nuclear targets.
Eq.~(\ref{eq:FA}) neglects the large $1<x<A$ tail in nuclei which
is (small but) nonzero due to Fermi motion.

\subsection{Comparison with the Collinear Parton Model}
\label{sec:coll}

So far, no reference to the parton model was required in
deriving the results. It is interesting to compare the above results, 
derived from the operator product expansion, with the collinear 
(${\vec k}_\perp = {\vec 0}$) parton model
approach to target mass corrections \cite{aot,kr}.
The latter also leads to Nachtmann scaling but the mixing
of the tensor basis comes in only from the non-collinear 
$k^- \neq \xi P^-$ 
component of the parton momentum. This approximation avoids
single and double-convolution terms of the form of Eqs. 
(\ref{eq:h1})-(\ref{eq:g2}).   
If the collinear approximation turns out to be of reasonable
accuracy, one may hope to employ similarly approximate mass
corrections to processes where the rigorous operator approach is not
available such as processes where the final state is not fully inclusive. 

\subsubsection{The CC Process including Charm Production}

We now find relations 
between $f_i$ and  ${\cal {F}}_{i}$, 
the parton model structure functions
for scattering off of the Cabibbo-Kobayashi-Maskawa-rotated weak
eigenstates (e.g., $d'=|V_{d,u}|^2 d+|V_{s,u}|^2 s$), normalized
such that for $\nu$ scattering at leading order in QCD, 
${\cal F}_i=(1-\delta_{i4})(d'+...)$. The full expressions for
${\cal F}_i$ appear in, for example, Eq.~(10) of Ref.~\cite{kr}.
To include charm production, we introduce the parameter
$\lambda\equiv Q^2/(Q^2+m_c^2)$ 
(i.e.~one should simply set $\lambda \rightarrow 1$ for 
the light quark contributions).
Our result is that
\begin{eqnarray}
f_{1,3,5}(\xi,Q^2)&&={2{\cal F}_{1,3,5}(\xi,Q^2)\over \xi}\\
f_2(\xi,Q^2)&&={2{\cal F}_2(\xi,Q^2)\over \lambda \xi}\\
f_4(\xi,Q^2)&&={4{\cal F}_4(\xi,Q^2)\over \xi}.
\end{eqnarray}
To make contact with 
the measurable structure functions $F_i$,
we instead write our results in terms of the partonic functions
${\cal F}_i$ and convolutions:
\begin{equation}\label{eq:FcalTMC}
F_j^{\rm TMC}=\sum_{i=1,5} \alpha_j^i {\cal F}_i(\xi,Q^2) + \beta_j^i 
{\cal H}_i(\xi,Q^2) + \gamma_j {\cal G}_2(\xi,Q^2)\ .
\end{equation}
The quantities ${\cal H}_i$ and ${\cal G}_i$ are 
\begin{eqnarray}
{\cal H}_i(\xi,Q^2) &&=\int_\xi^1 d\xi '{ {\cal F}_i(\xi ',Q^2)\over \xi'}\\
{\cal G}_i(\xi,Q^2) &&=\int_\xi^1 d\xi ' {\cal H}_i(\xi ',Q^2)
\end{eqnarray}
The coefficients $\alpha^i_j,\ \beta^i_j$ 
and $\gamma_j$ are listed in Tables \ref{tab:al}-\ref{tab:ga}. 

As can be seen from Table \ref{tab:al},
the $\alpha_j^i$ coefficients are mostly diagonal, however, there is some mixing
of the parton functions ${\cal F}_i$ for $F_4$ and $F_5$. This was already
seen in our analysis that included target mass effects in the
collinear limit. In fact the terms $\sum_i \alpha_j^i {\cal F}_j$
contributions to $F_i$ in Eq.~({\ref{eq:FcalTMC})
are a factor of $1/(1+\mu\xi^2)$ suppressed relative to our earlier
results Eqs.~(14)-(18) in Ref.~\cite{kr}
which included target mass corrections in the collinear limit.
The $k_T$ corrections are also
responsible for the remaining terms proportional to ${\cal H}_j$ and
${\cal G}_2$ in Eq.~(\ref{eq:FcalTMC}). 

\subsubsection{Massless Quark Limit and the NC Process}
\label{sec:ml}

Finite charm mass effects are mostly important at intermediate energies.
For low energy processes, charm production is a negligible effect, 
whereas at very high energies, charm can be treated effectively massless.
In these limits, we can evaluate the target mass corrections 
assuming three or four massless quarks. 
Only three independent helicity amplitudes survive the 
limit of massless quarks, or more generally the limit where the 
quark fields that build the currents in Eq.~(\ref{eq:wmunu0}) are
mass-degenerate.
This is discussed
in the appendix of Ref.~\cite{kr}. For massless quarks in CC interaction, 
or generally for the NC interaction we then have:
\begin{eqnarray}
{\cal F}_5&&={\cal F}_2\\
{\cal F}_4&&={1\over 2}({\cal F}_2-{\cal F}_1)
\end{eqnarray}
so $f_5=f_2$ and $f_4=f_2-f_1$.
Given these relations, one finds that 
\begin{eqnarray}
W_4&&=-\mu W_1+{1\over 4x^2}W_2\\
W_5&&={1\over 2x} W_2
\end{eqnarray}
which lead, with the usual definitions, to the expressions
\begin{eqnarray}
F_4&&={1\over 2} \Biggl[ { F_2\over 2x } - F_1
\Biggr] \\
F_5&&={1\over 2x} F_2\ ,
\end{eqnarray}
where all structure functions are understood to be including the
target
mass corrections of Eq.~(\ref{eq:FTMC}). Thus, in the massless quark limit
for CC interactions,
or for the NC process, the
evaluation of $F_1$, $F_2$ and $F_3$ is sufficient to fully describe the
target mass corrections, even when lepton masses cannot be neglected 
compared to the incident neutrino energy.  

\section{Applications}
\label{sec:appl}
\subsection{Operator Product Expansion versus Collinear Parton Model}

We have already commented on the comparison between the $\xi$ scaling
corrections derived from the OPE and those obtained in the collinear
parton model in Section \ref{sec:coll}. In this section we parallel this
discussion with a numerical evaluation of the charged current neutrino
scattering structure functions $F_2$ and $F_3$ 
in both of these approaches. 

Figs.~\ref{fig:f21}-\ref{fig:f34} show results of 
$F_2$ and $F_3$ at $Q^2=1$ and 4
GeV$^2$. The figures \ref{fig:f21}-\ref{fig:f34}~($a$) show the absolute scale of the structure
functions, while Figs.~\ref{fig:f21}-\ref{fig:f34}~($b$) show ratios of structure functions. With
($a$) and~($b$) together, 
one can avoid overstressing kinematic regions with
large corrections which are, nevertheless, not relevant because the
value of the structure function is small.
The structure functions in the figures are evaluated using the GRV 
parton distribution functions \cite{grv98}, which have 
3 active flavors. Here and below, we work in 
the convention of three active flavors and in the 
$\overline{{\rm MS}}$ factorization scheme.

Fig.~\ref{fig:f21}~($a$) shows $F_2$ at $Q^2=1$ GeV$^2$ as a function of $x$. The
solid curve is our full result of target mass corrected NLO
($\xi$ scaling corrections derived from the OPE) $F_2(\xi ,1\ 
\ {\rm GeV}^2)$.
The dot-dashed curve  is the NLO corrected $F_2^{(0)}(x, 1\ {\rm  GeV}^2)$
without TMC, and the dashed curve is the LO $F_2^{(0)}$.

In Fig.~\ref{fig:f21}~($b$), we show ratios of $F_2$'s at $Q^2=1$ GeV$^2$, all as a
function of $x$. 
The solid line shows the ratio of 
the LO $F_2(\xi, 1 \ {\rm GeV}^2)$ evaluated with the full TMC to
the collinear approximated TMC (TMC-COL) corrected $F_2$. The
collinear
approximation is discussed in Ref. \cite{kr} and is summarized by
\begin{equation}
F_j = \sum_{i=1,...,5} \alpha_j^i (2-\xi/x){\cal F}_i(\xi,Q^2).
\end{equation}
The dot-dashed line is the ratio of the full TMC corrected $F_2$ to
the TMC-COL $F_2$, both at NLO. 
One finds that the collinear approximation does reasonably well in
representing the full TMC structure function $F_2$. 

The dashed line in Fig~\ref{fig:f21}~($b$) is the ratio of the NLO TMC $F_2$ to the
NLO $F_2$ in the $M\rightarrow 0$ limit ($F_2^{(0)}$). The TMC are of
order 15\%  at $x=0.4$ and very large above $x\sim 0.7$. The region
$x>0.7$ is where the NLO $F_2^{(0)}$ is less than 10\%  of its value
at $x=0.1$.

Figs.~\ref{fig:f24}~($a$) and \ref{fig:f24}~($b$) show the same quantities for $Q^2=4$ GeV$^2$.
Figs.~\ref{fig:f31}~($a$), ~\ref{fig:f31}~($b$) and ~\ref{fig:f34}~($a$), ~\ref{fig:f34}~($b$) 
are the corresponding results for $F_3$. 

The comparative results may be summarized as follows
\begin{itemize}
\item As is well known, target mass corrections become important at low $Q$
and they are suppressed at low $x$.
\item For $0<x<0.9$, the collinear approximation deviates from 
the exact OPE results by at most 10\% for $F_2$ and at most 20\% for $F_3$.
A more detailed examination of the adequacy of the collinear 
approximation for phenomenology will be left to future work.
\end{itemize}
As briefly mentioned above, the second point may be
reason for some optimism (though by no means a proof) 
that simple kinematic
rescalings provide the dominant hadron mass 
effects in other processes as well.
A crucial observation here is that a rescaling term is enhanced by a 
derivative factor
\begin{equation}
F(\xi ) - F(x) = \frac{1}{2} \left. \frac{d F}{d \xi}
\right|_{\xi =x} (\xi -x) + ...
= {\cal{O}}\left( \frac{M^2}{Q^2} \right) F^{\prime} (x)
\end{equation}
for a steeply falling function $F(\xi )$. E.g.~for a toy function 
$F(x) \propto (1-x)^\beta $ the rescaling for $x \rightarrow 1$
is enhanced as $1/(1-x)$ compared to a multiplicative
$[1+{\cal{O}}(\frac{M^2}{Q^2})] F(x)$ correction term.

\subsection{The Paschos-Wolfenstein Relation and Related Observables}
\label{sec:PW}
In the following, integrated cross sections 
\begin{equation}\label{eq:sigint}
\sigma_{\rm NC, CC}^{\nu, {\bar \nu}}
= \frac{
\int d E_{\nu , {\bar \nu}}\ d \sigma_{\rm NC, CC}^{\nu, {\bar \nu}}
\ \Phi (E_{\nu {\bar \nu}})
\ \mid_{ 20\ {\rm GeV} < y E_{\nu , {\bar \nu}} < 180\ {\rm GeV}}
}
{
\int d E_{\nu , {\bar \nu}}\ \Phi (E_{\nu {\bar \nu}})
}
\end{equation}
will refer to flux-averaged integrals with a cut on hadronic energy as in
the experimental analysis \cite{nuanom}. We use the 
(anti-)neutrino fluxes in 
a step function form as shown in Fig~\ref{fig:flux}.
We will consider the {\it counting experiment} observables
\begin{equation} \label{eq:rnu}
R^{\nu,{\bar \nu}} \equiv 
\frac{\sigma^{\nu,{\bar \nu}}_{\rm NC}}{\sigma^{\nu,{\bar \nu}}_{\rm CC}}
\end{equation}
as well as the Paschos-Wolfenstein \cite{pw}
relation
\begin{equation}
\label{eq:pw}
R^- \equiv 
\frac{\sigma^{\nu}_{\rm NC}-\sigma^{\bar \nu}_{\rm NC} }
{\sigma^{\nu}_{\rm CC}-\sigma^{\bar \nu}_{\rm CC} }
\simeq \frac{1}{2} - \sin^2 \Theta_{\rm W}
\end{equation}
The question to what extent the approximation in Eq.~(\ref{eq:pw}) holds
in the light of the high precision data of Ref.~\cite{nuanom} has been of some debate
recently. For an ideally iso-scalar target and under the neglect of charm production
components one has that $R^- = 1/2 - \sin^2 \Theta_{\rm W}$ 
at arbitrary order in QCD as long as there 
are no $(s -{\bar{s}})(x)$ component
in the nucleon strange sea \cite{forte}. For definiteness, we 
will consider the scattering of neutrinos on an $Z=26, A=56$ iron target and 
charm production with $m_c=1.3\ {\rm GeV}$. We will be using
parton distributions \cite{grv98,ctq6} from the literature, though, which
do not contain a $(s-{\bar s})(x)$ asymmetry. In a somewhat idealized Mellin
moment language, NLO corrections have already been estimated in Ref.~\cite{forte}
and we refer the reader to this reference\footnote{While completing this article
the effect of NNLO moments were presented in Ref.~\cite{moch}.} 
for more detailed background information.
Our evaluation of Eqs.~(\ref{eq:rnu}), (\ref{eq:pw}) complements Ref. \cite{forte} in that
we do not make the approximations required for a moment analysis and that we 
mimic the experimental setup to some extent by including the neutrino flux and hadronic
energy cut in Eq.~(\ref{eq:sigint}). Technically, Eq.~(\ref{eq:sigint}) is evaluated as a
5-dimensional Monte Carlo integral:
\begin{equation} \label{eq:intdim}
\int d \sigma
\  d E_{\nu , {\bar \nu}}\propto \int d x\ d y\ d E_{\nu , {\bar \nu}}\ d \xi_{1,2}
\end{equation}
over the kinematic variables and two convolution variables arising from
NLO and TM correction integrals. Eq.~(\ref{eq:intdim}) integrates numerically
over plus type distributions in the perturbative coefficient functions. In a 
MC integral, this has to be done with care to achieve the required accuracy.

Our results for $R^{\nu , {\bar \nu}}$ are summarized in Table \ref{tab:rnu} for 
two sets (GRV \cite{grv98} and CTEQ6 \cite{ctq6}) of 
parton distribution functions, for a LO\footnote{
The LO evaluation also neglects target mass effects. The charm mass is kept
no-zero along the LO slow rescaling prescription.
} 
or NLO evaluation and for
the standard model and the anomalous value of the Weinberg angle.
The CTEQ6 parton distribution analysis also
provides a master formula -- Eq.~(3) in \cite{ctq6} --
to estimate the PDF related error on a physical observable
through a set of error PDFs.
Our results can be 
summarized as follows
\begin{itemize}
\item $R^{\bar \nu}$ is insensitive to the Weinberg angle and sensitive to NLO
corrections
\item $R^{\nu}$ is insensitive to NLO corrections within its sensitivity to
the Weinberg angle
\item The impact of PDF uncertainties from pre-determined PDF fits is inconclusive,
e.g.~the error estimate of CTEQ6 for $R^{\nu}$ does not overlap with the evaluation
based on GRV. This fact is not too surprising, though, given there are other factors
such as scheme (number of active flavours) uncertainties and that the error estimate
has not been tailored to match this particular high precision application.
\end{itemize}
From these results, one cannot derive a conclusive estimate of the impact of NLO corrections 
on the analysis \cite{nuanom} and further work will be required. 
For now, we will restrict ourselves to
playing the game to treat Eq.~(\ref{eq:pw}) as a would-be identity and to solve it
for the Weinberg angle:
\begin{eqnarray}\label{eq:l1}
\left. \frac{1}{2}-R^- \right|_{\{{\rm GRV\ LO},\ \sin^2 \Theta_{\rm W}=0.2227\}}
&=& 0.2192(3)\\ \label{eq:l2}
\left. \frac{1}{2}-R^- \right|_{\{{\rm GRV\ NLO},\ \sin^2 \Theta_{\rm W}=0.2227\}}
&=& 0.2192(2) \\ \label{eq:l3}
\left. \frac{1}{2}-R^- \right|_{\{{\rm CTEQ\ NLO},\ \sin^2 \Theta_{\rm W}=0.2227\}}
&=& 0.2196(9) \pm 0.0005(1)\ .
\end{eqnarray}
The numbers in parentheses are the next digit in the numerical
evaluation. 
For the evaluation of $R^-$, in addition to the values for $R^{\nu,{\bar \nu}}$
in Table \ref{tab:rnu} one also needs the values for
\begin{equation}\label{eq:r}
r \equiv 
\frac{\sigma^{\bar \nu}_{\rm CC}}
{\sigma^{\nu}_{\rm CC}}
\end{equation}
in Table \ref{tab:r} giving
\begin{equation}
R^- = \frac{R^{\nu} - r R^{\bar \nu}}{1-r}\ .
\end{equation}

The difference between the numerical values in Eqs.~(\ref{eq:l1}) and Eqs.~(\ref{eq:l2})
reflects the impact of a LO or NLO evaluation of the cross sections entering $R^-$.
It is actually beyond the numerical precision of our calculation which is about $10^{-4}$. 
The error quoted with the NLO evaluation using CTEQ6M refers to the master formula (3) in
Ref.~\cite{ctq6} and has to be understood as explained in detail in this reference.
For the observable $R^-$ we find a very robust stability under NLO corrections even 
under the non-ideal conditions of a non-isoscalar target and in a world of massive
charm quarks and a massive target. We also find a similar stability with regards to 
PDF variations as long as they do not exploit
any new physical degree of freedom such as isospin
violations or an asymmetry $(s-{\bar{s}})(x)$ in the 
strange sea.
From theory alone we cannot investigate if the
correlations between $R^{\nu}$ and $R^{\bar \nu}$ in the analysis \cite{nuanom} are
correctly represented by the combined ratio $R^-$. The values in 
Eqs.~(\ref{eq:l1}) - (\ref{eq:l3}), nevertheless, support the expectation 
that the NuTeV anomaly is not a 
technical NLO effect.

\section{Summary}
\label{sec:sum}

In oscillation searches, neutrino cross sections relate event rates with neutrino fluxes
or the absence of events with bounds on oscillation parameters. 
A recent measurement of the Weinberg angle which deviates from the standard model
expectation is based on the observation of neutrino interactions on iron.
Contrary to the photon exchange case, the massive boson propagator is non-divergent
at vanishing invariant boson mass. Neutrino cross sections are, therefore, dominated
by perturbative interactions with quarks and gluons provided the neutrino 
energy is high enough. The dominance of deep inelastic interactions typically sets in
already above a few GeV neutrino energy. 
Typical $Q^2$ values are not high enough, though,
to rely on the lowest order parton model picture for 
anything but a crude estimate. Subleading power suppressed terms and logarithmic 
corrections are unlikely to be separable in the data and
have to be considered on the same footing in theory. 
In this article we have revisited, summarized 
and extended the inclusion of perturbative NLO 
corrections and heavy quark as well as lepton and target mass terms
into the twist-2 component of weak structure functions and perturbative 
neutrino cross sections. Further corrections not considered here are then, 
by definition, either of NNLO perturbative accuracy or of higher twist. 
A factor which we have barely discussed here and which does not
seem to lend itself to a systematic treatment easily
is the impact of our imperfect knowledge of parton distribution functions.
Obviously, this is even more true for higher twist parton correlations.
This article may be seen as a technical basis and
starting point to address some of these points
in more detail in the future. We have set up a Monte Carlo integration program for
weak structure functions and integrated cross sections and, for now, provided
numerical results for two warm-up applications: First, we have investigated the
accuracy of target mass corrections in the approximate collinear parton model which
we found to be better than 10\% for $F_2$ and better than 20\% for $F_3$. 
This result is interesting for computational efficiency
and as an expectation for the accuracy of
similar approximations to processes where the OPE is not
applicable. Second, we have looked into observables that are related to the 
measurement of the weak mixing angle in neutrino scattering. Though we confirmed
the expected stability of the Paschos-Wolfenstein relation, 
our first results do not allow any final conclusion on the impact of corrective
terms to an experimental analysis based on a different cross section model.
Future work will be done along this direction which will also take PDF 
uncertainties into account.

\section*{Note Added on Moment Estimates}
After we submitted this article, in addition to 
Refs.~\cite{forte,moch}
analytic moment estimates of NLO corrections to neutrino cross sections 
were also presented in Ref.~\cite{dobell}.
The authors point out a different sign for the NLO correction to $R^\nu$
than we find. 
Under the same approximations\footnote{The approximations required for
making moment estimates and obtaining energy-independent ratios
$R^{\nu,{\bar\nu}}$ are: $(i)$ neglect of evolution, $(ii)$ 
$M_{Z,W}\rightarrow \infty$ limit of the boson propagator,
$(iii)$ neglect of the hadronic energy cut.} 
and conventions\footnote{I.e., to reproduce Ref.~\cite{dobell} we
switch to the 4 flavor DIS factorization scheme ($m_c=0$)
adopted there, instead of our
canonical 3 flavor ${\overline{\rm MS}}$ convention ($m_c\neq 0$). 
} 
as used in \cite{dobell}, we reproduce the 
results in Eqs.~(5.14) and (5.15) of that paper.
We have then taken the approximate moment approach  
as a starting point and examined the implications of the different
approximations it makes. 
First, we find that the sign of the NLO correction to $R^\nu$
is not flipped by the hadronic
energy cut \cite{nuanom} in Eq.~(\ref{eq:sigint}). The hadronic energy cut
was not included in \cite{dobell}. 
Some of the other approximations made 
in \cite{dobell}, however, can flip the sign of the correction
to $R^\nu$ and thus explain the
seeming discrepancy between our calculation and that of \cite{dobell}.

An approximation that has implications for the sign of the NLO correction
to $R^\nu$ is the moment approximation itself.
The exact $\mu$ dependence of the parton distribution
functions
and coefficient functions is parametrically of order
$\alpha_s^2({\overline {Q^2}})
\ln (Q^2 / {\overline {Q^2}})$ (for average $Q^2\equiv \overline{Q^2}$) 
but its neglect overestimates (underestimates) typical $Q$ values
at low $x$ (large $x$) where structure 
functions are larger (smaller) for increasing $Q$. 
This leads to an overestimate of neutrino cross-sections 
of about
$\sim 10\%-15\%$ around $E_\nu = 100\ {\rm GeV}$. In addition,
for fixed parton distributions (or their second moments), there is
a factorization scheme dependence in the NLO correction 
and a comparison of different conventions
is of limited physical significance. 
Also, the classification of LO and NLO terms is generally not the
same for $n_f=3$ or $n_f=4$ active parton flavours.
We find that both the $Q \simeq {\overline Q}$ 
approximation and the dependence on the factorization scheme 
conventions are capable of flipping the sign of the NLO correction to
$R^\nu$.
More relevant of course than the volatile sign of this correction is
the question -- posed in \cite{dobell} and above 
in Section \ref{sec:PW} --  of what impact a complete and consistent 
QCD treatment of neutrino cross sections has on the analysis
of the data in Ref.~\cite{nuanom}. More detailed numerical results 
related to NuTeV observables will be presented in a 
future publication \cite{kr3}.

\section*{acknowledgments}
We thank B.~Dobrescu, K.~Ellis, K.~McFarland, W.~Marciano, S.~Moch, F.~Olness, 
W.-K.~Tung and G.~Zeller for discussions and correspondence 
and G.~Zeller also for providing us
with tabulated values of the NuTeV neutrino fluxes. 
S.K.~is grateful to RIKEN, Brookhaven National Laboratory and
the U.S.~Department of Energy (contract No.~DE-AC02-98CH10886)
for providing the facilities essential for the completion of this work.
The work is also supported by the U.S.~Department of Energy 
under Contract No.~FG02-91ER40664. M.H.R. thanks the Aspen Center
for Physics for its hospitality.

%%%%%%%%%%%%%%%%%%%%%%%%%%%%%%%%%%%%%%%%%%%%%%%%%%%%%%%%%%%%%%
\newpage
\setcounter{equation}{0}
\def\theequation{A\arabic{equation}}
\section*{Appendix A: General DIS cross sections}
\label{sec:app}

The electroweak weights in Eq.~(\ref{eq:sigdis}) for a 
general $(v_l-a_l)$ leptonic current are given by
\begin{eqnarray} \label{eq:wa1}
w_1 &=& 
\frac{E_{l} y \left[(m_1^2+4 m_2 m_1 +m_2^2 + 2 E_{l} M x y)a_l^2
+v_l^2 (m_1^2-4 m_2 m_1 +m_2^2 + 2E_l M x y)\right]}
{4 M (E_l^2 -m_1^2)}\\ \label{eq:wa2}
w_2 &=& -\frac{\left[4(y-1)E_l^2+2 M x y E_l+(m_1+m_2)^2\right]a_l^2+
v_l^2\left[ 4(y-1)E_l^2+ 2 M x y E_l+(m_1-m_2)^2\right]}
{8 (E_l^2-m_1^2)}\\ \label{eq:wa3}
w_3 &=& -\frac{a_l v_l E_l \left[-m_1^2+m_2^2+2 E_l M x(y-2)\right]y}
{4 M (E_l^2 -m_1^2)}\\ \label{eq:wa4}
w_4 &=& 
\frac{a_l^2\left[(m_1-m_2)^2+2 E_l M x y\right](m_1+m_2)^2 +
v_l^2 (m_1-m_2)^2\left[ (m_1+m_2)^2+2 E_l M x y\right]}
{8 M^2 (E_l^2-m_1^2)x}
\\ \label{eq:wa5}
w_5 &=& 
 -\frac{E_l\left[(m_1+m_2)(m_2+m_1(y-1))a_l^2+v_l^2(m_1-m_2)(y m_1-m_1-m_2)
\right]}
{2 M(E_l^2-m_1^2)}
\end{eqnarray}
For CC neutrino scattering ($v_l=a_l=1$, $m_1=0$) this reduces to
\begin{eqnarray} \label{eq:wb1}
w_1 &=& 
\frac{m_2^2 y}{2 M E_l} + x y^2 \\ \label{eq:wb2}
w_2 &=&
1-\frac{m_2^2}{4 E_l^2}-\left(1+\frac{M x}{2 E_l}\right)y\\ \label{eq:wb3}
w_3 &=& 
-\frac{m_2^2 y}{4 M E_l} +\frac{x}{2}(2-y)y\\ \label{eq:wb4}
w_4 &=& 
\frac{m_2^4}{4 E_l^2 M^2 x}+\frac{m_2^2 y}{2 M E_l} \\ \label{eq:wb5}
w_5 &=& 
-\frac{m_2^2}{M E_l}
\end{eqnarray}
In the above Eqs.~(\ref{eq:wb1})-(\ref{eq:wb5})
one has $m_2=m_{\tau}$ for tau neutrinos and
for electron or muon neutrinos $m_2=m_{e,\mu}\simeq 0$.
For NC interactions ($m_1=m_2=m$) we have 
\begin{eqnarray} \label{eq:wc1}
w_1 &=& 
\frac{E_l y \left[m^2(3a_l^2-v_l^2)+E_l M(a_l^2+v_l^2)x y\right]}
{2 (E_l^2-m^2)M}\\ \label{eq:wc2}
w_2 &=& 
-\frac{E_l v_l^2\left[ 2E_l (y-1)+x y M\right]
+a_l^2\left[2 m^2+2E_l^2(y-1)+E_l M x y\right]}
{4 (E_l^2-m^2)}
\\ \label{eq:wc3}
w_3 &=& 
-\frac{v_l a_l E_l^2 x(y-2)y}{2 (E_l^2-m^2)}
\\ \label{eq:wc4}
w_4 &=& 
\frac{a_l^2 E_l m^2 y}{M(E_l^2-m^2)} 
\\ \label{eq:wc5}
w_5 &=& 
-\frac{a_l^2 E_l m^2 y}{M(E_l^2-m^2)}
\end{eqnarray}
where $m=0$
for neutrino neutral current scattering. For charged lepton 
electromagnetic interactions the $w_i$  
in Eqs.~(\ref{eq:wc1})-(\ref{eq:wc5}) with  
$\{a_l,w_{i>2}\} = 0$, $v_l = 1$ 
have to be inserted into Eq.~(\ref{eq:sigdisem}).

%%%%%%%%%%%%%%%%%%%%%%%%%%%%%%%%%%%%%

\begin{table}
\caption{Coefficients $A_j^i$ in Eq.~(\ref{eq:FTMC}).\label{tab:A}}
%\begin{ruledtabular}
\begin{tabular}{lccccc}
$A_j^i$ & $i=1$          & $i=2$ & $i=3$ & $i=4$ & $i=5$\\
$j=1$ & ${x\over \xi\rho}$ & 0     & 0     &   0   &    0 \\
$j=2$ & 0              &${x^2\over \rho^3\xi^2}$ & 0 & 0 & 0\\
$j=3$ & 0 & 0 & ${ x\over \rho^2\xi}$ & 0 & 0\\
$j=4$ & 0 & ${\mu^2 x^3}\over \rho^3$ & 0 & 
${1 \over (1+\mu\xi^2)}$ &
$-{2\mu x^2 \over \rho^2}$\\
$j=5$ & 0 &$ -{\mu x^2\over \rho^3 \xi}$ & 0 & 0 & ${ x\over \rho^2\xi}$\\
\end{tabular}
%\end{ruledtabular}
\end{table}

\begin{table}
\caption{Coefficients $B_j^i$ in Eq.~(\ref{eq:FTMC}). \label{tab:B}}
%\begin{ruledtabular}
\begin{tabular}{lccccc}
$B_j^i$ & $i=1$          & $i=2$ & $i=3$ & $i=4$ & $i=5$\\
$j=1$ & 0 & ${\mu x^2\over \rho^2}$      & 0     &   0   &    0 \\
$j=2$ & 0              &${6 \mu x^3\over \rho^4}$ & 0 & 0 & 0\\
$j=3$ & 0 & 0 & ${2\mu x^2
\over \rho^3}$ & 0 & 0\\
$j=4$ & 0 & ${-2\mu^2  x^4\over \rho^4}(2-\mu\xi^2)$ & 0 & 0 &
${\mu x^2\over \rho^3}$\\
$j=5$ & 0 &$ {2\mu x^2 (1-\mu\xi x)\over \rho^4}$ & 0 & 0 & ${\mu x^2
\over \rho^3}$\\
\end{tabular}
%\end{ruledtabular}
\end{table}

\begin{table}
\caption{Coefficients $C_j$ in Eq.~(\ref{eq:FTMC}).\label{tab:C}}
%\begin{ruledtabular}
\begin{tabular}{lcccc}
$C_1$ & $C_2$          & $C_3$ & $C_4$ & $C_5$ \\
${2\mu^2 x^3\over \rho^3}$ & ${12\mu^2 x^4\over \rho^5}$ & 0 & 
${2\mu^2 x^3\over \rho^5}(1-2\mu x^2)$ & ${6 \mu^2 x^3\over \rho^5}$ \\
\end{tabular}
%\end{ruledtabular}
\end{table}
%%%%%%%%%%%%%%%%%%%%%%%%%%%%%%%%%%%%%%%%%%%%%%%%%%%%%%%%%%%%%%%%%%%%%%%%%%%%

\newpage
\begin{table}
\caption{Coefficients $\alpha_j^i$ in Eq.~(\ref{eq:FcalTMC}).\label{tab:al}}
%\begin{ruledtabular}
\begin{tabular}{lccccc}
$\alpha_j^i$ & $i=1$          & $i=2$ & $i=3$ & $i=4$ & $i=5$\\
$j=1$ & ${x\over \xi\rho}$ & 0     & 0     &   0   &    0 \\
$j=2$ & 0              &${2 x^2\over\lambda \rho^3\xi}$ & 0 & 0 & 0\\
$j=3$ & 0 & 0 & 2${ x\over \rho^2\xi}$ & 0 & 0\\
$j=4$ & 0 & ${2\mu^2 \xi^2 x^2\over\lambda \rho^2 (1+\mu\xi^2)}$ & 0 & 
${1\over (1+\mu\xi^2)}$ &
$-{2\mu x \xi\over \rho(1+\mu\xi^2)}$\\
$j=5$ & 0 &$ -{2\mu x^2\over\lambda \rho^3}$ & 0 & 0 & ${ x\over \rho^2\xi}$\\
\end{tabular}
%\end{ruledtabular}
\end{table}

\begin{table}
\caption{Coefficients $\beta_j^i$ in Eq.~(\ref{eq:FcalTMC}).\label{tab:be}}
%\begin{ruledtabular}
\begin{tabular}{lccccc}
$\beta_j^i$ & $i=1$          & $i=2$ & $i=3$ & $i=4$ & $i=5$\\
$j=1$ & 0 & ${2 \mu x^2\over\lambda \rho^2}$      & 0     &   0   &    0 \\
$j=2$ & 0              &${12\mu x^3\over\lambda \rho^4}$ & 0 & 0 & 0\\
$j=3$ & 0 & 0 & ${4\mu x^2
\over \rho^3}$ & 0 & 0\\
$j=4$ & 0 & ${-4\mu^2  x^4\over\lambda \rho^4}(2-\mu\xi^2)$ & 0 & 0 &
${2\mu x^2\over \rho^3}$\\
$j=5$ & 0 &$ {4\mu x^2 (1-\mu\xi x)\over\lambda \rho^4}$ & 0 & 0 & ${2\mu x^2
\over \rho^3}$\\
\end{tabular}
%\end{ruledtabular}
\end{table}

\begin{table}
\caption{Coefficients $\gamma_j$ in Eq.~(\ref{eq:FcalTMC}).\label{tab:ga}}
%\begin{ruledtabular}
\begin{tabular}{lcccc}
$\gamma_1$ & $\gamma_2$          & $\gamma_3$ & $\gamma_4$ & $\gamma_5$ \\
${4\mu^2 x^3\over \lambda \rho^3}$ & ${24\mu^2 x^4\over\lambda \rho^5}$ & 0 & 
${4\mu^2 x^3\over\lambda \rho^5}(1-2\mu x^2)$ & ${12 \mu^2 x^3\over\lambda \rho^5}$ \\
\end{tabular}
%\end{ruledtabular}
\end{table}

%%%%%%%%%%%%%%%%%%%%%%%%%%%%%%%%%%%%%%%%%%%%%%%%%%%%%%%%%%%%%%%%%%%%%%%%%%%%
\newpage
\begin{center}

\begin{minipage}[t]{10cm}
\begin{table}
\begin{tabular}{l|l|l}
PDF ($\sin^2 \Theta_{\rm W}$)  &  $R^\nu$  &  $R^{\bar \nu}$ \\
\hline 
GRV NLO (0.2227) & 0.3120 (0.3120) & 0.3844 (0.3845)\\
GRV LO (0.2227)  & 0.3125 & 0.3860 \\
GRV NLO (0.2277) & 0.3088 & 0.3839 \\
CTEQ6 NLO (0.2227) & $0.3105 \pm 0.0006$ & 0.$3841 \pm 0.0038$
\end{tabular}
\caption{The ratios 
$R^{\nu, {\bar \nu}}$ 
as defined in 
Eq.~(\ref{eq:rnu})
evaluated for different parton distributions 
{\protect\cite{grv98,ctq6}}, 
at leading and higher order 
and for two values of the Weinberg angle. The error quoted for the CTEQ6 PDF refers
to the master formula (3) in 
{\protect\cite{ctq6}} 
and must be understood as explained in
detail in this reference. In the first line, the numbers in parentheses refer to
a perturbative expansion of the ratios $R^{\nu ,{\bar \nu}}$ directly
(instead of the ratios of perturbatively expanded
cross sections in Eq.~(\ref{eq:rnu})); i.e.~schematically
$R^\nu = R^\nu_{(0)}+\alpha_s R^\nu_{(1)}$ inside the parentheses.
\label{tab:rnu}}
\end{table}

\begin{table}
\begin{tabular}{l|l}
PDF &  $r$  \\
\hline 
GRV NLO & 0.3009 \\
GRV LO & 0.3014  \\
CTEQ6 NLO & 0.2909
\end{tabular}
\caption{The ratio $r$ 
as defined in 
Eq.~(\ref{eq:r})
evaluated for different parton distributions 
{\protect\cite{grv98,ctq6}}, 
at leading and higher order. 
\label{tab:r}}
\end{table}
\end{minipage}
\end{center}

%%%%%%%%%%%%%%%%%%%%%%%%%%%%%%%%%%%%%%%%%%%
\begin{center}

\begin{figure}
\psfig{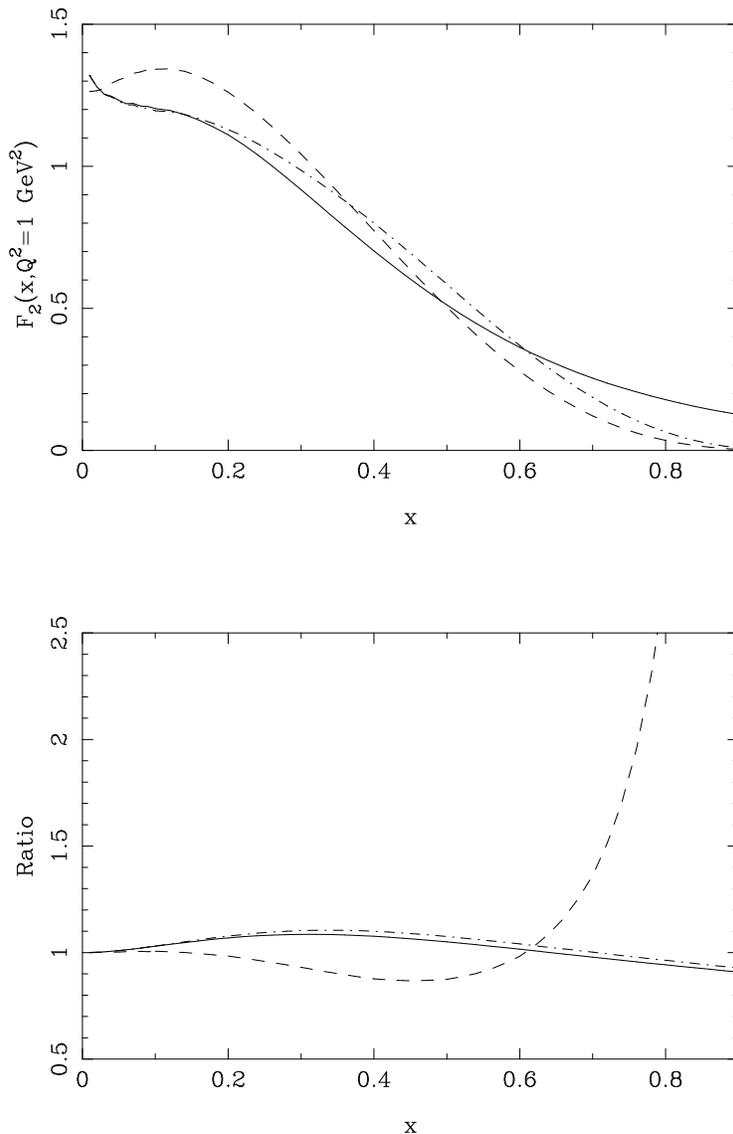}
\caption{($a$) The charged current neutrino structure function $F_2$ at
$Q^2=1\ {\rm GeV}^2$ evaluated with NLO $\xi$-scaling corrections
(solid line)
and in LO (dashed) and NLO (dot-dashed)
under the neglect of target mass corrections.
($b$) Ratio of the LO $\xi$-scaling evaluation of
$F_2$ as in ($a$) with $Q^2=1$ GeV$^2$
to the LO evaluation in the
collinear approximation is shown by the solid line.
The dot-dashed line shows the same ratio  at NLO. The dashed line
shows the ratio of the NLO target mass corrected $F_2$ to the NLO
$F_2$ in the $M\rightarrow 0$ limit.
\label{fig:f21}
}
\end{figure}

\begin{figure}
\psfig{figure=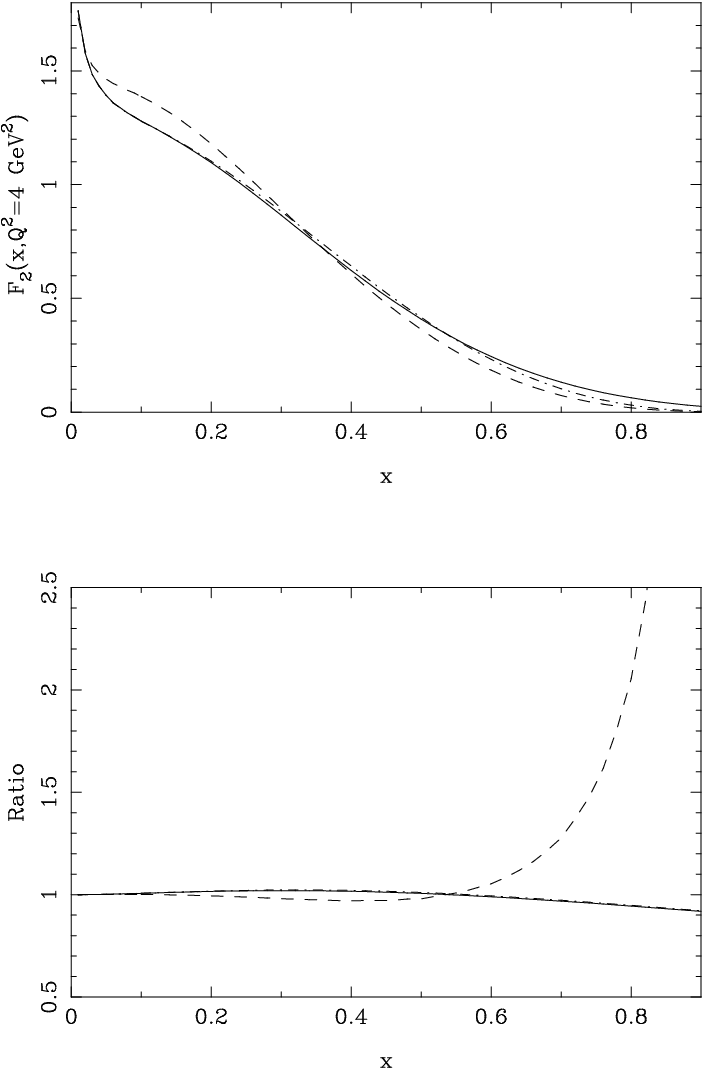,width=15cm,angle=270}
\caption{Same as Fig.~\ref{fig:f21} for $Q^2=4\ {\rm GeV}^2$.
\label{fig:f24}
}
\end{figure}

\begin{figure}
\psfig{figure=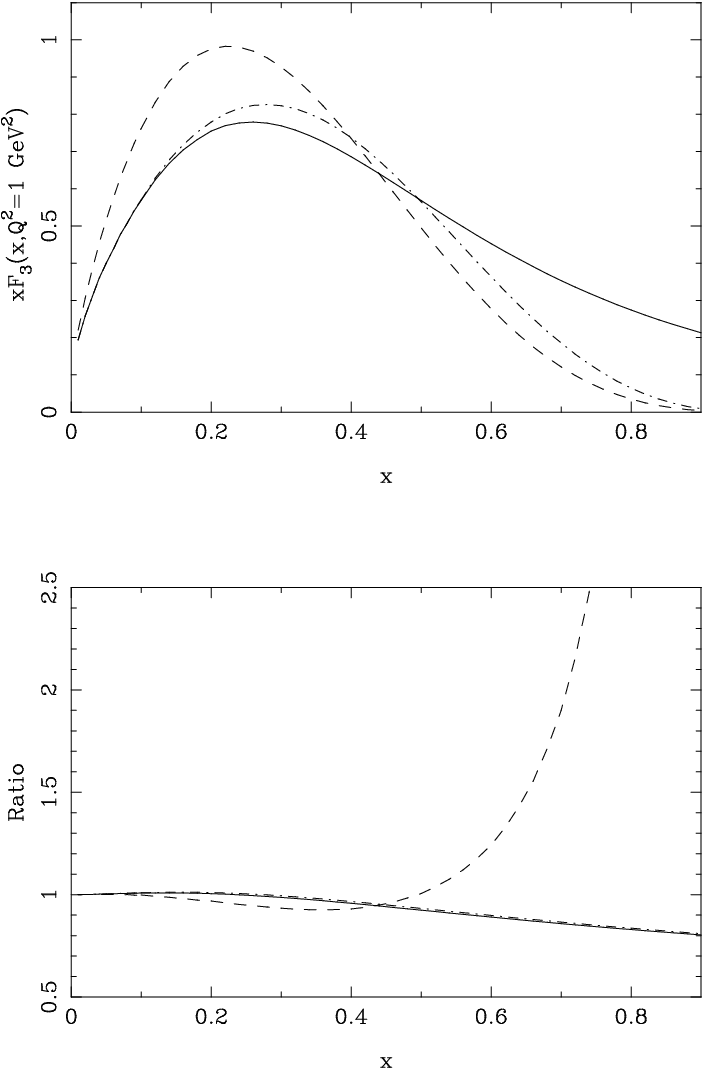,width=15cm,angle=270}
\caption{Same as Fig.~\ref{fig:f21} for the structure function
$xF_3$.
\label{fig:f31}
}
\end{figure}

\begin{figure}
\psfig{figure=xf3q24.ps,width=15cm,angle=270}
\caption{Same as Fig.~\ref{fig:f31} for $Q^2=4\ {\rm GeV}^2$.
\label{fig:f34}
}
\end{figure}

\begin{figure}
\psfig{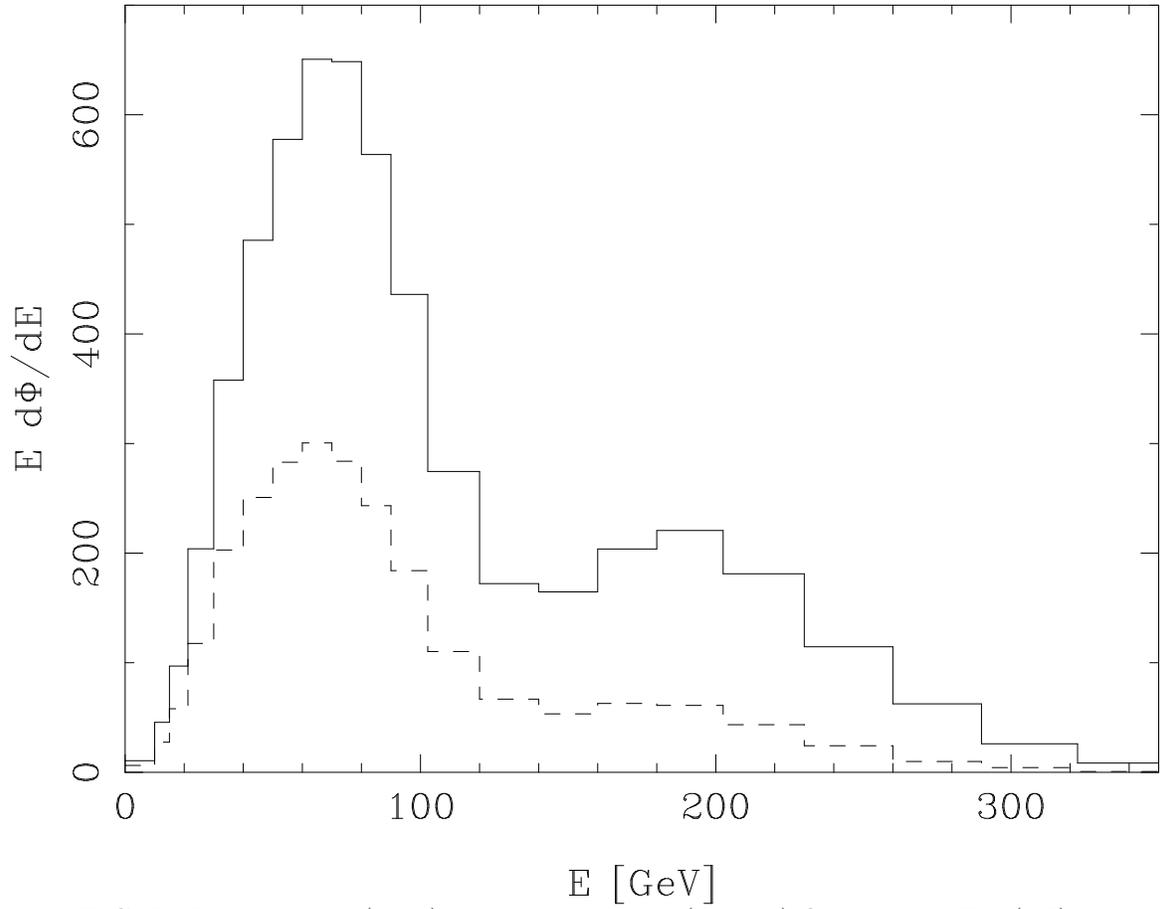}
\caption{The neutrino (solid) and anti-neutrino (dashed)
flux used in Eq.~(\ref{eq:sigint}).
\label{fig:flux}
}
\end{figure}

\end{center}

%%%%%%%%%%%%%%%%%%%%%%%%%%%%%%%%%%%%%%%%%%%%%%%%%%%%%%%%%%%%%

\end{document}